\documentclass[12pt]{article}
\usepackage{graphicx}
\usepackage{color}
\usepackage{amsmath,slashed, amssymb}
\usepackage{fancyhdr}
\usepackage{hyperref}
\usepackage[titletoc]{appendix}
\numberwithin{equation}{section} 

\def\beq{\begin{eqnarray}}
\def\eeq{\end{eqnarray}}
\def\bea{\begin{eqnarray*}}
\def\eea{\end{eqnarray*}}

\def\centeron#1#2{{\setbox0=\hbox{#1}\setbox1=\hbox{#2}\ifdim
\wd1>\wd0\kern.5\wd1\kern-.5\wd0\fi
\copy0\kern-.5\wd0\kern-.5\wd1\copy1\ifdim\wd0>\wd1
\kern.5\wd0\kern-.5\wd1\fi}}
\def\ltap{\;\centeron{\raise.35ex\hbox{$<$}}{\lower.65ex\hbox{$\sim$}}\;}
\def\gtap{\;\centeron{\raise.35ex\hbox{$>$}}{\lower.65ex\hbox{$\sim$}}\;}

\def\singleandthirdspaced{\baselineskip=\normalbaselineskip\multiply
    \baselineskip by 130\divide\baselineskip by 100}

\newcommand{\newc}{\newcommand}
\newc{\qbar}{{\overline q}}
\newc{\Kahler}{Kahler }
\newc{\deltaGS}{\delta_{\rm GS}}
\begin{document}
\begin{titlepage}
\begin{flushright}
{\large SCIPP 15/04\\UTTG-08-15 \\
}
\end{flushright}

\vskip 1.2cm

\begin{center}

{\LARGE\bf Tunneling in Theories with Many Fields}

\vskip 1.4cm

{\large Michael Dine$^{(a)}$, Sonia Paban$^{(b),(c)}$}
\\
\vskip 0.4cm
{\it $^{(a)}$Santa Cruz Institute for Particle Physics and
\\ Department of Physics, University of California at Santa Cruz \\
     Santa Cruz CA 95064  } \\
\vspace{0.3cm}
{\it
$^{(b)}$Department of Physics and Texas Cosmology Center \\ University of Texas at Austin\\ Austin,TX 78712}
\\  \vspace{0.3cm}
{\it
$^{(c)}$Institute for Advanced Study \\ Princeton, NJ 08540}
\\     \vspace{0.3cm}

\vskip 4pt

\vskip 1.5cm

\begin{abstract}
The possibility of a landscape of metastable vacua raises the question of what fraction of vacua are truly long lived.
Naively  any would-be vacuum state has many nearby decay paths, and all possible decays must be suppressed.
An interesting model of this phenomena consists of $N$ scalars with a random potential of fourth order.  Here we show
that the scaling of the typical minimal bounce action with $N$ is readily understood, and differs from statements in the literature.  We
discuss the extension to more realistic landscape models.
\end{abstract}

\end{center}

\vskip 1.0 cm

\end{titlepage}
\setcounter{footnote}{0} \setcounter{page}{2}
\setcounter{section}{0} \setcounter{subsection}{0}
\setcounter{subsubsection}{0}
\setcounter{figure}{0}

%%%%%%%%%%%%%%%%%%%%%%%%%%%%%%%%%%%%%%%%%%%
%%%%%%%%%%%%%%%%%%%%%%%%%%%%
\singleandthirdspaced

\section{Introduction:  The Existence of States in a Landscape}

The evidence for the existence of a landscape of vacua in theories (the theory?) of quantum gravity can at best be described as limited.
First there is the phenomenological fact of the peculiar value of the cosmological constant.   Second, there is the observation\cite{boussopolchinski,douglaskachru}
that it is possible to turn on large numbers of fluxes in some string compactifications, and these can take rather large values.  For a typical choice of fluxes, however,
systematic study is not possible, so one can't reliably determine whether or not there exist stable or metastable states and their properties.

In practice, most state counting is based on examining a potential for fields computed in a (not necessarily valid) semiclassical approximation, and
counting stationary points.   Apart from the question of whether any systematic approximation is available for the analysis,
Banks has argued\cite{bankseft} that this analysis does not make sense in a quantum theory of gravity. Without confronting
such difficult issues, the minimal questions one might ask about this procedure are:  what fraction of these points are local minima of the potential,
and of these, what fraction are sufficiently metastable to be compatible with a universe like ours.  The most interesting result which might emerge from such studies is the existence of special classes of solutions which exhibit larger fractions of (suitably metastable) ``vacuum states".

For the first question, suppose the would-be state in question contains $N$ scalar fields.  We require that the $N$ eigenvalues of the mass-squared matrix
are positive.  One's naive guess is that, without supersymmetry or some dynamical considerations, all eigenvalues are positive $(1/2)^N$ of the time.   In ref. \cite{mcallister}, however, modeling a landscape with a plausible random matrix model, it was shown that the suppression might be significantly more severe -- as $e^{-c~N^2}$,
with $c$ an order one constant.  One might hope that supersymmetry would lead to stability, since in general, in flat space with exact supersymmetry, all scalar masses are positive.  It is argued in \cite{mcallister} and \cite{Bachlechner:2014rqa}, however,  that there is still an exponential suppression with $N$ in this case.  We will comment on this issue later in this paper, expressing some skepticism but leaving the question for future research.

Refs. \cite{Dine:2007er,dfs} raised the question of quantum stability in a flux landscape.  With rare exceptions, a classically stable state of small c.c. would be surrounded
by many others of large, negative c.c. (other issues in landscape tunneling have been discussed in \cite{shiutunneling,tyetunneling} ). It was argued that in a flux landscape, the bounce actions would generally be small, scaling as an
inverse power of the typical flux, so there would
be many decay channels with unsuppressed amplitudes.   These authors asked what might account for a small tunneling amplitude to
every nearby state, and argued that supersymmetry broken at a scale well below
the fundamental scale was the most promising possibility.

The authors of \cite{greeneweinbergtunneling} put forward a field theory model in which to address the question
of quantum metastability (an interesting alternative model has been discussed, from a somewhat different viewpoint, in \cite{Dienes:2008qi}).  The model contains $N$ scalar fields.  The authors assume the existence of a classical
local minimum, and expand the potential in a power series about that point, choosing the coefficients
at random.   The authors
performed a numerical simulation, obtaining scaling laws for minimal barrier heights, distances in field space to the barrier, and 
bounce actions.
Their results indeed indicate that long-lived states are likely to be extremely rare in such a landscape.
The authors presented results on statistics of
distances to nearby saddle points and their heights.  In this note, we provide a simple explanation
of the results obtained for these features of the potential.  In particular, within the model, we argue that tunnelings in the direction of the smallest mass provide an upper bound on the barrier
height and the distance.   With a slight modification of the model, they permit a precise computation of the median
values of these quatities.  The authors of \cite{greeneweinbergtunneling} employed a proxy for the tunneling amplitude in terms of
the distances and heights, which lead to a particular scaling law for the tunneling amplitude.  Our analysis allows a more direct
computation, and leads to a different scaling law.

This understanding of the model, we will see, yields a close relation between the issues raised in
ref. \cite{mcallister} of classical stability and the problem of quantum metastability.   In a landscape model where one studies stationary points of some potential,
the question of metastability is that of the likelihood that the masses-squared of all scalars are positive.  For quantum metastability,
the requirement, as we will see, is (roughly) that all masses be larger than some number.  Depending on the nature of the landscape
and its potential, this can provide either modest or substantial further suppression of the number of states.

In the next section, we will present the model of \cite{greeneweinbergtunneling}, and derive the scaling laws for
mean lowest barrier heights
and distance to the barrier with $N$.  We will see that tunneling is typically dominated by the lightest scalars, and as a result the tunneling
amplitude is controlled by the mass and self-couplings of these scalars.  This permits, in a semiclassical limit, a calculation of the
tunneling amplitude.  As we explain, if the semiclassical analysis {\it is not} valid, then one expects that tunneling
is unacceptably rapid.  We are able to provide the scaling of the lowest bounce action with $N$.
In section \ref{connecting}, we discuss extending models like that of \cite{mcallister} to determine what fraction of
stationary points in a model landscape will exhibit quantum metastability.  We will see that, depending on the nature
of the underlying landscape, the additional suppression can be comparable to,
or more modest than, the requirements of classical metastability.  In section \ref{implications}, we conclude with a discussion of
implications of these results.  We recall the analysis which indicates that states with some degree of supersymmetry exhibit
quantum metastability.  We explain that, while these results do not rule out  the possibility of non-supersymmetric states in a landscape,
they do suggest that supersymmetry might be more {\it generic} than naively expected.
We note that, even if supersymmetric states predominate, these arguments by themselves do not necessarily suggest that supersymmetry is broken at a low energy (TeV or multi TeV) scale; merely that it should be broken at scales well below the Planck scale or other relevant fundamental scales.  In an appendix,
we explain the probability analysis.

\section{A Model Landscape}
\label{modellandscape}

The authors of ref. \cite{greeneweinbergtunneling} put forth a simple model for a landscape:  a theory with $N$ scalar fields, $\phi_i$, interacting
through a potential $V(\phi_i)$.  They imagine
that they are studying a particular stationary point of the potential, defined to be $\phi_i =0$, which is classically stable.  They
expand the potential about that point, keeping cubic and quartic terms:
\beq
V =  \left (\sum_i  \mu^2_i \phi_i^2 + \sum_{ijk} \gamma_{ijk} \phi_i \phi_j \phi_k + \sum_{ijkl} \lambda_{ijkl} \phi_i \phi_j \phi_k \phi_\ell
\right )
\eeq
Our notation differs somewhat from that of \cite{greeneweinbergtunneling} but the content is the same.  The parameters $\mu_i^2$, $\gamma_{ijk}$
and $\lambda_{ijkl}$ are taken to be random variables, with ranges, following \cite{greeneweinbergtunneling}:
\beq
0 < \mu_i^2 < M^2;  ~~-M < \gamma_{ijk} < M;~~-1< \lambda_{ijk} < 1.
\eeq
$M$ is some fixed mass which will scale out of our problem.  It is easy to modify the results we obtain below at large $N$ for different ranges of the parameters\footnote{The authors of \cite{greeneweinbergtunneling} also introduced a parameter $\lambda$ multiplying the potential.  By rescaling the parameters, this can be absorbed into the ranges of $\mu_i^2, \gamma_{ijk}$, etc.}.  Note that the vacua are assumed classically metastable.

The question which interests us is what fraction of vacua are significantly metastable.   Assuming the validity of a semiclassical
analysis, this is the question:  what fraction of states have bounce action for all possible bounce solutions greater than some fixed, large
number $B_0$? This assumption is self consistent.  If all tunneling amplitudes are small, all bounce actions are large,
justifying the semiclassical analysis.  Following \cite{greeneweinbergtunneling}, we will ignore gravitational effects, commenting on them in the concluding section.

In \cite{greeneweinbergtunneling}, computer simulations of this problem were reported.  The authors searched for stationary points of the action, looking for nearby critical points with low barriers.  They then applied a crude model for the bounce action.  They were able to perform their analysis for as many as $10$ fields.  They found:
\begin{enumerate}
\item  The distance to the nearest stationary point behaves as
\beq
\phi_{top} \approx 0.5 N^{-1.15}:
\eeq
\item  The height of the lowest stationary point behaves as:
\beq
V_{top} \approx 0.2 N^{-3.16}
\eeq
\item  The lowest bounce action scales as:
\beq
B \approx N^{-2.7}.
\eeq
\end{enumerate}
These scalings are similar for both cubic and quartic potentials.

In \cite{pabantunneling}, it was noted that some aspects of these results could be understood with certain assumptions about the distributions of quartic couplings.  Here we point out that
the first two of these results can be understood by simple statistical reasoning.  This analysis leads to a quite different scaling of the overall
bounce action than found in \cite{greeneweinbergtunneling}.

What is striking about the results above is the rapid decrease with $N$, and the similar behavior for cubic and
quartic potentials.  One might hope to provide a simple explanation of these scalings.
To this end, we consider first straight line trajectories in the field space.  In some directions the potential will grow indefinitely; these are not 
interesting for tunneling.  In others, the potential will turn over, becoming negative for a while or indefinitely. One possible explanation for the 
scaling with $N$ would be rapid growth of the effective cubic and quartic couplings for large $N$. 
Ref. \cite{pabantunneling,aravindtunneling} noted that if the quartic couplings grew like $N^2$ and the cubic like $N^{3/2}$, one could account for these scaling
laws.  But simply thinking in terms of random walks such growth is hard to explain.

But there is a simpler explanation, which immediately provides
bounds on these quantities compatible with the observed numerical results.
For large $N$, one has some small masses, $\mu_i^2$ of order $1/N$.  Let's proceed first under the
assumption that the smallest bounce action (and lowest barrier and shortest distance to tunnel) are obtained
in one of these directions.     Call $i=1$ the direction with smallest $\mu^2$.  Let's assume, first, that the lowest bounce action
is obtained by a straight line trajectory in the $1$ direction.  The important cubic and quartic couplings are then
$\gamma_{111} \equiv \gamma$, $\lambda_{1111}\equiv \lambda$ and these will typically be of order $1$.
In this case, the cubic term dominates, and
\beq
\phi_{top} = {2 \over 3} {\mu^2 \over \gamma}~~~~V_{top} = -{4 \over 27} \left ({\mu^6 \over \gamma^2} \right ),
\eeq
with corrections of order $1/N$.  

Even with our assumption of straight line trajectories along particular mass eigenstates,
the assumption that the lightest scalar dominates is not exactly correct.  The cubic and quartic couplings, in particular, can fluctuate downward, in which case one of the larger
masses may dominate.
A more careful analysis gives, for the median values of $\phi_{top}$, for large $N$:
\beq
\phi_{top} =  .924 ~N^{-1}
\eeq
and
\beq
V_{top} = 0.284 N^{-3}.
\eeq
This is compatible with the results of \cite{greeneweinbergtunneling}, up to corrections of order $1/N$.  These authors worked to $N=10$.

Our real interest is in the scaling of the tunneling amplitude.  Here ref. \cite{greeneweinbergtunneling} makes a crude approximation.
In a manner reminiscent of the thin-wall approximation, the authors take the bounce action to be
\beq
B = {\pi^2 \over 2} \sigma R^3
\eeq
where $\sigma$ is essentially a one dimensional bounce action and $R$ is the bubble radius, both estimated by considering straight line paths to the nearest
saddle point of the potential.  Given the $N$ scaling of the barrier height and width, 
\beq
\sigma \propto \int d\phi \sqrt{V} \sim N^{-5/2}
\eeq
roughly as they find.

However, knowing that the tunneling trajectories are dominated by small $\mu^2$, we can do a more systematic
calculation of the large $N$ tunneling behavior.  Given the domination by the cubic term, we are interested in tunneling
in a potential of the form
\beq
V = \mu^2 \phi^2 - \gamma \phi^3
\eeq
Simple scaling arguments give
\beq
\phi(r) = \frac{\mu^2}{\gamma} \phi_0(r\mu),
\eeq
where $\phi_0$ is the bounce for the potential $V= \phi^2-\phi^3$, and the bounce action scales as $\mu^2/\gamma^2$. 
Sarid\cite{sarid} has studied this problem numerically, obtaining
\beq
B = 2.376 \times 2 \pi^2 {\mu^2 \over \gamma^2} \label{approxaction}.
\eeq
One sees immediately that, for large $N$, the lowest bounce action typically behaves as $\pi^2/N$.  More careful analysis
gives, for the median bounce action at large $N$:
\beq
B_{med} = {97.5 \over N} 
\eeq
So if $N= 100$, for example, the exponential of the typical bounce action is not large.

Note that if we allow the ranges of parameters to be:
\beq
\mu_i^2 < a_1 M^2;~~\vert \gamma_{ijk} \vert  < a_2 ~M~~~\vert \lambda_{ijkl} \vert < a_3
\eeq
and, as in \cite{greeneweinbergtunneling} we include an overall factor of $\lambda$ in the potential, $B$, our expression for the median bounce action becomes 
\beq
B_{med} = {a_1 \over \lambda a_2^2}{97.5 \over N}.
\eeq

We are actually interested in  the probability
that the lowest action satisfies, say
\beq
B > B_0 \eeq
for some constant $B_0$.
For actions of the form (\ref{approxaction}), the probability distribution will be (\ref{pdfW}):
\beq
P(B>B_0)= P( w>w_0) = \left\{ \begin{array}{cc}  w_0<1 & \left(1-\frac{w_0}{3} \right)^N \\  \\ w_0>1 & \left( \frac{2}{3 \sqrt{w_0}} \right)^N \end{array} \right.
\label{wprobability}
\eeq where $B \equiv  2.376  \times 2 \pi^2 w$.  Requiring that the $B_0$ give a lifetime for the most rapid tunneling process
longer than the age of the universe gives $w_0 >5.7$.  To get some feeling
for numbers, taking $N=100$, this is a suppression of order 
$10^{-56}$.

We need, however, to reexamine our assumption that the straight line trajectory defined by a scalar field of definite mass
yields the bounce with the lowest action.  Quite generally, we expect to be able to lower the bounce action, at least slightly, by
simply considering straight line trajectories in slightly different directions in the field space, and by studying paths which are not
straight lines.
But the real worry is that there are directions which effectively have large $\gamma$.  If we consider a field direction,
$\Phi$, with
\beq
\phi_i = a_i \Phi + \dots~~~\sum a_i^2 = 1
\eeq
then in this direction, the quadratic and cubic couplings are
\beq
\mu^2 = \mu_i^2 a_i^2; ~~~\Gamma = \gamma_{ijk} a_i a_j a_k.
\eeq
If the typical (median) maximum $\Gamma$ grows with $N$ as $N^p$, then, given that the typical $\mu^2$ will be of order $1/2$,
the distance to the nearest minimum will behave as $N^{-p}$, and the barrier height will behave as $N^{-2p}$.  The bounce action
will behave as $N^{-2p}$.  So if $p =1/2$, this direction will be competitive with the direction of the smallest mass.  If
$p>1/2$, it will dominate.

To assess this, we have studied numerically the value of the maximum $\Gamma$ for randomly chosen $\gamma_{ijk}$ for $N$
as large as $40$.  We find no evidence for growth of the (median) maximum $\Gamma$ with $N$, much less the growth
required for this to provide the dominant trajectory.  At the same time, for modest $N$, this direction {\it is} competitive
with the smallest mass direction.  For the full range of $N$, we find that the median $\Gamma$ is $2.2$, so $\mu^2/\Gamma^2 \approx 0.1$.  This should be compared to $2/N$ from the small mass directions.  So small mass ``wins" only for $N> 20$ or so.

%Suppose $N$ is large, and consider the $k$ scalars with the smallest masses.
%To understand the issues, start with $k=2$, and suppose $\gamma_{211} \ne 0$.  Then turning on a small value of $\phi_2$,
%one can lower the height and distance to the barrier.  Similarly we can lower the bounce action.  Thought in terms of directions
%in field space, this corresponds to rotating the original direction to a direction with a component along $\phi_2$.  As one makes the 
%angle larger, however, the mass also becomes larger, and so one expects, typically, to find an order one decrease of the
%bounce action.  If one picks $k$ fields, one has a larger space to explore, but the range of $\mu^2$'s will be of order $k/N$.
%So one might expect a similar, ${\cal O}(1)$ suppression.    We have checked this numerically, by seeking the minimum, with
%several fields, of the tunneling trajectory (and barrier heights) along straight line trajectories.  We find that the suppression is
%by a factor only slightly larger than $1$.

But a realistic landscape, if such exists, might not resemble the model of \cite{greeneweinbergtunneling}.  In particular, there might be correlations among various couplings, and one might expect that there would be some sort of locality in the space of field (indices).  So,
for example, one might guess that
\beq
\gamma_{ijk} = {a_{ijk} \over 1 + A \left ( (i-j)^2 + (j-k)^2 + (i-k)^2 \right )}
\eeq
with $a_{ijk}$ a similarly distributed set of random numbers, would provide a more realistic model.  In this case, the typical
$\Gamma$ would be small (and there would not grow with $N$.  Similar correlations among the elements of the mass
matrix would have implications for the suppression found in \cite{mcallister}.)

%In a more realistic model with large numbers of fields, we might expect that many of the analogs of the $\gamma_{i11}$ vanish, further
%mitigating this effect.
%\beq
%P(B>\hat{B})= P( w>\hat{w}) = \left\{ \begin{array}{cc}   \hat{w}<1 & \left(1-\frac{\hat{w}}{3} \right)^N \\  \\ \hat{w}>1 & \left( \frac{2}{3 \sqrt{\hat{w}}} \right)^N \end{array} \right.
%\eeq where $B \equiv  2.376  \times 2 \pi^2 w$.

\section{Connecting Classical and Quantum Stability}
\label{connecting}

We have focussed here on quantum
stability, but he first question one might explore is the likelihood that all mass-squareds are positive in a non-supersymmetric
or nearly supersymmetric theory.  Actually, our results in the previous section suggest that these problems are similar.
Naively as we have noted, one might expect that with $N$ fields,
one would have a suppression of $2^N$ of the probability that a given stationary point is in fact a minimum.  If one requires,
instead, that all masses be larger than some given number, and all $\gamma$'s suitably small, 
one has an additional suppression, as we have seen.

But the suppression might be much stronger.
We have mentioned the work of \cite{mcallister} on the distribution of eigenvalues of the scalar mass matrix in
supergravity theories.  Here the focus was on classical stability, i.e. the probability of obtaining all
eigenvalues of the mass-squared matrix positive.  The authors worked in the framework
set out in \cite{douglasdenef}.  That work studied the distribution of SUSY breaking scales around
stationary points of a supergravity potential motivated by Type II string theories compactified with fluxes.  Among other 
results, they found a distribution heavily concentrated at the highest possible supersymmetry breaking scales.

The analysis of \cite{mcallister}, as in \cite{douglasdenef},
was conducted in the framework of a supergravity effective action, keeping only terms with two derivatives, and modeling
particular terms in the mass matrix with suitable random matrices.
 The principal observation was that the lowest eigenvalue is typically negative, so one wants
the probability for a fluctuation in which all eigenvalues are positive. 
Within the particular random matrix model, these authors found that (classical) metastability, in the absence of approximate supersymmetry,
was highly suppressed, as $e^{-c N^2}$.  With approximate supersymmetry, they found suppression as $e^{-c^\prime N^{1}}$.

%Having recognized the importance of 
%the mass matrix for quantum metastability, the question of the distribution of these eigenvalues takes on additional
%relevance.   Where one might  the suppression goes as $e^{-c~N^2}$, for some constant $c$.  Among states with some approximate supersymmetry, suppression of stability is not quite so extreme.

One might question the use of a supersymmetric action given the dominance of large supersymmetry breaking.  If $F$ (the decay constant of the Goldstino superfield) is of order $M_p$,  there is no reason to ignore terms with arbitrary numbers of (covariant) derivatives.  So perhaps an approach with weaker assumptions is to consider $N$ real fields and describe their mass matrix in terms of symmetric matrices.  For these, there is a reasonably well-developed theory of fluctuations, and one can write, for the probability that all masses are positive\cite{Dean:2006wk}:
\beq
P = e^{-N^2 {\ln(3) \over 4} + {\cal O}(N)}.
\eeq
This is compatible with the results of \cite{mcallister}.

But there is reason to be skeptical about this  extreme suppression.
We have commented in the previous section on the possibility of correlations between various terms in the action.
We suggested there that one might expect some degree of locality within the space of field indices.  If this is the case, then
one would expect something closer to the naive result, $P = e^{-aN}$ for the probability that all masses are positive.  Of course, this would still be an enormous
suppression, but, unlike the more extreme case, one might imagine that there are still significant numbers of states for large $N$.

%If the mass matrix is sparse, for example if it breaks up into $k \times k$ blocks%
%(where $k$ is not proportional to $N$),
%the suppression might not be so severe, behaving as $e^{-a N}$ for some constant $a$.
%Before considering the possible implications of these observations for a landscape, we turn to the question of non-perturbative
%stability, i.e. stability against tunneling.  

In our tunneling discussion, we are focussed on precisely the small subset of states which are not merely classically stable but
for which all of the masses, $\mu_i^2$, are positive and larger than some fixed number (more precisely that $\mu^2/\gamma^2$
larger than some number).  In the language 
of ref. \cite{mcallister}, the question is one of finding a fluctuation where {\it all} masses fluctuate to some value larger than some
positive $\mu_0^2$.  If we imagined a flat, uncorrelated, distribution (now running, from, say, $-1$ to $1$, the ``naive" estimate above)
we would expect the a suppression similar to that we obtained from eqn. \ref{wprobability} over and above that required to obtain
all masses positive.
%\beq
%P = \left ({1 \over 2} \right )^N \rightarrow \left ({3 \over 4} \right )^N ~~~ e^{-0.7N} \rightarrow e^{-1.4 N}
%\eeq
%a huge suppression on top of an already huge suppression.

%In our tunneling discussion, we are focussed on precisely the small subset of states which are not merely classically stable but
%for which all of the masses, $\mu_i^2$ are positive and large.  In the language 
%of ref. \cite{mcallister}, the question is one of finding a fluctuation where {\it all} masses fluctuate to some value larger than some
%positive $\mu_0^2$.  If we imagined a flat, uncorrelated, distribution (now running, from, say, $-1$ to $1$, the ``naive" estimate above) we would, in other words,
%expect something like:
%\beq
%P = \left ({1 \over 2} \right )^N \rightarrow \left ({3 \over 4} \right )^N ~~~ e^{-0.7N} \rightarrow e^{-1.4 N}
%\eeq
%a huge suppression on top of an already huge suppression.
%In a model landscape like that of \cite{mcallister}, we might expect even stronger suppression.
In a model landscape like that of \cite{mcallister}, we might expect even stronger suppression.
Indeed, we can again consider the case of random symmetric matrices.  Our discussion of tunneling suggests that we require a further, strong constraint, that all masses-squared be larger than some fixed number.
If we assume, as before, that cubic and quartic couplings are roughly uniformly distributed random variables, with typical values of order the fundamental scale, then from our previous analysis, requiring all tunneling amplitudes to be larger than the age of the universe, we require:
\beq
{\mu^2 \over \gamma^2}  > 2.9.
\eeq
To assess this probability, we can again borrow results from \cite{Dean:2006wk}. These authors determined the probability that all of the eigenvalues (masses-squared for our problem) are larger than some value, $z$.  For $z \ll \sqrt{N}$, one has the additional suppression:
\beq
P(z) = \exp(-{2 {\sqrt{6} \over 9} z~N^{3/2}})
\eeq
while for $z \gg \sqrt{N}$,
\beq
P(z) = \exp(-54 N z^2).
\eeq

To estimate the size of $z$, we need to ask what is a typical $\gamma$ (we will not attempt a serious analysis including fluctuations
in $\gamma$).  We might imagine $\gamma \sim M_p$, supposing $M_p$ to be a typical scale.  In this case, $z \ll \sqrt{N}$, and
the additional suppression required by quantum stability, while substantial, is in some sense a minor correction to that required by
classical stability.  On the other hand, in flux landscapes, the couplings in the lagrangian scale like the {\it square} of a typical flux.
So the bound on $\mu^2$ also scales as the square, and this could well lead to an enormous additional suppression. 

\section{Implications of The Suppression}
\label{implications}

In modeling the possible existence of a landscape, one typically considers some structure, such as compactified string theories
with fluxes, and counts the number of stationary points of the resulting effective action.  This represents the ``state of the art", but
it raises many questions. For example, following \cite{greeneweinbergtunneling}, we have assumed the
validity of a semiclassical analysis, and neglected gravitational effects.  The neglect of gravitation is
reasonable if all of the relevant scales are small compared to the Planck scale, but we might expect that at typical stationary
points, all scales are comparable and dimensionless couplings are of order one. In such a regime, on the other hand,
there is no small parameter which might account for the long lifetime of would-be states. We would simply argue that the rarity of stability
in cases for which one does (might) have control supports the expectation that stability among strongly coupled/Planck scale states
is exceptional.

Confronted with the huge suppression of metastable states, both classically and quantum mechanically, it is natural to ask:
does there exist a landscape at all, in the sense of some vast number of states?  Conceivably this sort of analysis (combined
with arguments like those of \cite{dfs} about decays with flux emission) would leave only a modest number (i.e. not exponentially
large) of metastable states.  But it is also possible that there are simply so many stationary points that a dense landscape of states
exists.  Resolving the issue of whether the suppression is $e^{-aN^2}$ or $e^{-cN}$ may well be critical to answering this question. 

A different viewpoint on this result has to do with the possible role of supersymmetry in an underlying theory.
A longstanding question is the extent to which supersymmetry might be typical of states in a landscape.  One might expect that it is special, perhaps exponentially rare, and so unlikely to account for hierarchies.  This viewpoint was explored, for example, in \cite{douglassusy,susskindsusy}.  However, considerations of stability point in the opposite direction.  First, classically, if we require that the cosmological constant is small, unbroken or slightly
broken supersymmetry implies classical stability for all but a small subset of fields.  So one, at least naively, expects that an order one
fraction of these states will be classically stable.  References \cite{mcallister} and \cite{Bachlechner:2014rqa} found a quite different result, and we will 
comment on this in a moment. As stressed in \cite{dfs}, unbroken or slightly broken supersymmetry implies quantum
stability\footnote{Similar to the classical case, this does not consider the possibility of tunneling along directions involving a finite set
of light fields.  The question of further exponential suppression will be studied elsewhere.}.  Among stationary points, supersymmetry could be quite rare, and still overwhelm the enormous suppression
which we have discussed.

The question of possible exponential suppression with $N$ in the supersymmetric case is related to the question of correlations.
in the models studied in \cite{mcallister} and \cite{Bachlechner:2014rqa}, it was assumed that there is a Goldstino field, $Z$, along with matter fields, $\phi_i$,
with couplings
\beq
W = Z~f+ a_i Z^2 \phi_i + {m_i \over 2} \phi_i^2
\eeq
Integrating out the massive fields leads to a contribution to the $Z$ mass:
\beq
\delta m_Z^2 = - \sum_{i=1}^N \vert a_i \vert^2 \vert { f^2 \over 4m_i^2 }\vert.
\eeq
If the $N$ fields have comparable mass, and the $a_i$'s are all similar, then for large $N$, the probability that the $Z$ mass-squared is positive is exponentially small.  But one might well imagine that only a few of the $a_i$'s are substantial (in particular, a number which does not scale with $N$).  

Stability, by itself, is not an argument for TeV scale supersymmetry; it is presumably enough that the scale of supersymmetry breaking be a few orders of magnitude below the fundamental scale (Planck, string(?)).  Indeed, in \cite{branches}, it was argued that, with dynamical supersymmetry breaking but random value for the superpotential, the distribution of supersymmetry breaking scales was roughy constant, decade by decade.  Lower scales might arise if the superpotential itself was dynamical.  Our arguments don't address the relative likelihood of these various possibilities.  Stability, then, raises the troubling possibility that supersymmetry does exist at comparatively low energies, but not necessarily at scales which would be accessible to any conceivable accelerator.  Sharpening these arguments is clearly of great importance.

\begin{appendices}

\section{Random Variables}
In this section we will give the probability density functions (p.d.f) of several functions of random variables that appear in the paper. 

Let X be a uniformly distributed random variable, then its p.d.f. will be 

\beq
f_{X}(x)= \left\{ \begin{array}{cc}
1 & 0<x<1 \\
0 & \mbox{otherwise} \\
\end{array} \right. \label{pdfX}
\eeq
The p.d.f for a variable $Z \equiv X^2$, where $X$ is defined as above
\beq
f_{Z}(z)= \left\{ \begin{array}{cc}
\frac{1}{2 \sqrt{z}}  & 0<z<1 \\ \\
0 & \mbox{otherwise} \\
\end{array} \right. \label{pdfZ}
\eeq
The p.d.f of a variable $Y\equiv X/Z$, where $X$ and $Z$ have  p.d.f. defined in (\ref{pdfX}) and (\ref{pdfZ}) respectively, is:
\beq
f_{Y}(y)= \left\{ \begin{array}{cc}
\frac{1}{3}  & 0<z<1 \\ \\
\frac{1}{3 z^{3/2}} & z>1 \\
\end{array} \right. \label{pdfY}
\eeq
Finally, let $W= min\{ Y_j, j=1, \cdots, N \}$, where all the $Y_j$ are distributed according to (\ref{pdfY}), its p.d.f. will be 

\beq
f_{W}(w)= \left\{ \begin{array}{cc}
N \left( 1- \frac{w}{3} \right)^{N-1}  & w<1 \\ \\
\frac{N}{3 w^{3/2}} \left( \frac{2}{3 \sqrt{w} }\right)^{N-1} &w>1\\
\end{array} \right. \label{pdfW}
\eeq
For large $N$ the median of (\ref{pdfW}) is given by 

$$ w_M= \frac{3 \ln 2}{N} + O(N^{-2}) $$

\end{appendices}

\vspace{1cm}

\noindent
{\bf Acknowledgements:} SP thanks T. Bachlechner, B. Greene, A. Masoumi, L. McAllister and E. Weinberg for useful discussions and A. Barvinok and P. Vivo for correspondence.   M.D. and S.P. thank the Institute for Advanced Study
for hospitality and support during the course of this investigation.  M.D. also thanks
the Technion Israel Institute of Technology for hospitality.  This work was supported by the U.S. Department of Energy grant number DE-FG02-04ER4128 and by a grant from the Simons Foundation ($\#305975$ to Sonia Paban). 

\bibliographystyle{unsrt}
%\bibliographystyle{JHEP}
%\bibliography{Biblio}
%\bibliography{dinerefs}{}
\bibliography{multifield_tunneling_jhep.bbl}{}
%\bibliography{tunnelingrefs}{}
%\bibliographystyle{utphys}
%\bibliographystyle{unsrt}
%\bibliographystyle{JHEP}
%\bibliography{Biblio}

\end{document}